\documentclass[11pt]{article}
\usepackage{latexsym,cite}
\usepackage{graphicx}% Include figure files
\usepackage{dcolumn}% Align table columns on decimal point
\usepackage{bm}% bold math
\usepackage{xcolor}
\usepackage{float}
\usepackage{epsfig}
\input amssym.def
\input amssym.tex

\newcommand{\be}{\begin{equation}}
\newcommand{\ee}{\end{equation} }
\newcommand{\beqa}{\begin{eqnarray}}
\newcommand{\eeqa}{\end{eqnarray} }
\newcommand{\ba}{\begin{array}}
\newcommand{\ea}{\end{array}}

\newcommand\cN{{\cal N}}

\newcommand\cZ{{\cal Z}}

\newcommand\M{{\Delta}}
\newcommand\rmd{{\rm d}}

\newcommand\state{{\psi}}

\newcommand\kB{k_{{\scriptscriptstyle{\rm B}}}}
\newcommand\leff{l_{{\scriptscriptstyle{\rm eff.}}}}
\newcommand\LdeB{\Lambda_{{\scriptscriptstyle{\rm de\,Broglie}}}}

\newcommand\fst{1}
\newcommand{\red}[1]{{\color{red} #1 \color{black}}}
\newcommand{\green}[1]{{\color{green} #1 \color{black}}}
\newcommand{\blue}[1]{{\color{blue} #1 \color{black}}}
\newcommand{\orange}[1]{{\color{orange} #1 \color{black}}}
\newcommand\superc{{{\scriptscriptstyle{\rm supercool}}}}
\newcommand\superh{{{\scriptscriptstyle{\rm superheat}}}}

\begin{document}
\begin{titlepage}
\title{\vskip -140pt
%\vskip 20pt
\vskip 2cm  Thermodynamic instability and first-order\\ phase transition in an ideal Bose gas\\~\\}
\author{\sc Jeong-Hyuck Park${}^{\dagger\ast}$ and  Sang-Woo Kim${}^{\natural}$}
\date{}
\maketitle \vspace{-1.0cm}
\begin{center}
~~~\\
${}^{\dagger}$Department of Physics $\&$ Center for Quantum  Spacetime\\Sogang University,  Mapo-gu, Seoul 121-742, Korea\\
%\texttt{park@sogang.ac.kr}
~{}\\
${}^{\natural}$High Energy Accelerator Research Organization (KEK)\\ Tsukuba, Ibaraki 305-0801, Japan\\ 
%Center for Quantum Spacetime, Sogang University\\ Shinsu-dong, Mapo-gu, Seoul 121-742, Korea\\
~{}\\
~~~\\
\end{center}
\begin{abstract}
\vskip0.1cm
\noindent
We conduct a rigorous  investigation into  the thermodynamic instability of ideal Bose gas confined in a cubic box, without assuming thermodynamic limit  nor  continuous approximation.  Based on the   exact expression of   canonical partition function, we perform  numerical computations up to the number of particles one million. We report that   if the number of particles is equal to or greater than a certain critical  value, which turns out to be ${7616}$, the ideal Bose gas  subject to  Dirichlet boundary condition    reveals a thermodynamic instability. 
Accordingly     we   demonstrate - for the first time -  that, a system consisting of  finite number of particles  can  exhibit   a discontinuous  phase transition featuring   a  genuine mathematical singularity, provided we keep not volume but pressure constant. The specific   number, ${7616}$   can  be regarded   as a characteristic number of `cube'  that   is     the geometric shape of the box. 
\end{abstract}
%%%
{\small
\begin{flushleft}
~~~~~~~~\textit{PACS}: 03.75.Hh,  05.70.Fh, 12.40.Ee, 51.30.+i\\
%%%
%% 03.75.Hh Static properties of condensates; thermodynamical, statistical, and structural properties 
%% 05.70.Fh Phase transitions: general studies  
%% 12.40.Ee Statistical models  
%% 51.30.+i Thermodynamic properties, equations of state  
%%%
~~~~~~~~\textit{Keywords}: thermodynamic instability, first-order phase transition, \\
~~~~~~~~~~~~~~~~~~~~~~~ideal Bose gas.\\
~~~~~~~~${}^{\ast}$\textit{Corresponding electronic address}: park@sogang.ac.kr
\end{flushleft}}
%%%
\thispagestyle{empty}
%%%%
\end{titlepage}
\newpage

\section{Introduction}
By definition,   first-order phase transitions in thermodynamics  feature    a
genuine mathematical singularity. Whether finite systems in Nature  can  literally exhibit such an infinity is a long standing  controversial question~\cite{Felderhof,Kastner}.

When a thermodynamic  system is composed  of  a definite  number of particles, say $N$, and  is in contact with a heat reservoir, the key quantity is  the canonical partition function:
\be
\textstyle{Z_{N}(\beta,V)=\sum_{\state}\, e^{-\beta E_{\state}}\,,}
\label{ZNgeneral}
\ee
where  the sum is over all the  quantum states of  the $N$-body system.  When the energy eigenvalues depend on the volume of the system, $Z_{N}$ is a function of the volume $V$, and the temperature through $\beta=1/(\kB T)$, where $\kB$ denotes Boltzmann constant. Provided  the precise  canonical partition function, we may compute various physical quantities, which include the pressure, the entropy, the internal energy  and the specific heat per particle at constant volume as follows:
\be
\ba{ll}
P=(1/\beta)\partial_{V}\ln Z_{N}\,,~&~~S=\kB(1-\beta\partial_{\beta})\ln Z_{N}\,,\\
E=-\partial_{\beta}\ln Z_{N}\,,~&~~C_{V}=(\kB/N)\beta^{2}\partial^{2}_{\beta}\ln Z_{N}\,.
\ea
\label{PSEC}
\ee
In particular, $\partial^{2}_{\beta}\ln Z_{N}=\langle(E_{\state}-E)^{2}\rangle$ being a  standard deviation squared, $C_{V}$ is finite and never negative. Further, the temperature derivative of  the probability for the system to occupy a  certain quantum state $\state$  reads
\be
{\textstyle{\left.\frac{\partial~}{\partial T}\right|_{V}
\left(e^{-\beta E_{\state}}/Z_{N}\right)=\kB\beta^{2}
\left(E_{\state}-E\right)e^{-\beta E_{\state}}/Z_{N}}\,.}
\label{TProb}
\ee
As the temperature increases from absolute zero to infinity,  the corresponding probability increases  if the  energy eigenvalue is greater than the average  \textit{i.e.} $E_{\state}>E$, and  it  starts to decrease in the opposite case,  $E_{\state}<E$.

Since the canonical partition function appears as a positive definite analytical function of its arguments,  clearly all the   physical quantities listed  in (\ref{PSEC}) and (\ref{TProb})  should not feature any singularities.  They may do so only in the thermodynamic limit: the limit of $N\rightarrow\infty$ and $V\rightarrow\infty$ with $N/V$ held fixed~\cite{LeeYang}.

However, strictly speaking,  infinite  limits   are  hardly realistic and exist only in theory~\cite{Felderhof,Kastner,LebowitzRMP,MoreSame}.   The above analysis seems to suggest  that Nature does not admit   a  discontinuous  phase transition featuring a genuine mathematical singularity,  which is somewhat different from  experiments or our daily experiences under the standard pressure 1 atm.

 In this paper we pay attention to, among others, the fact that  the above finiteness and continuity are for the cases  of keeping  the volume fixed. Once we switch to an alternative constraint of  keeping the pressure constant, we demonstrate,  for the first time,  that canonical ensembles with finite number of physical degrees may undergo   a discontinuous  phase transition.     \\

 The organization  of the present paper is as follows:\\
 Section \ref{sectionGENERAL} is devoted to a systematic analysis on the thermodynamic instability of a generic finite system. We explain how   canonical ensembles with finite number of physical degrees may exhibit  a discontinuous  phase transition when we keep not volume but pressure constant. Further we present  a \textit{theorem} which states  that,  a thermodynamic  system must be  unstable at low temperature near  absolute zero if the ground state energy is volume independent.  Examples include systems with vanishing ground state energy, such as supersymmetric models, ideal Bose or Boltzmann gases  under periodic or Neumann boundary conditions.

 In  section \ref{sectionIDEAL},  as a concrete model we focus on ideal Bose gas which is confined in a box and subject to Dirichlet boundary condition. Since there is no zero mode, the ground state energy depends on the volume and our \textit{theorem} is  not applicable in this case.  Nevertheless, by numerical analysis we show that the ideal Bose gas  reveals a thermodynamic instability and consequently  undergoes   a first-order phase transition,  if the number of particles is equal to or greater than ${7616}$.

 The final section \ref{sectionDISCUSSION} conveys our discussion.  In particular, we comment on the similarity between the permutation symmetry of the identical particle indices and the gauge symmetry in  high energy physics.

 Appendix contains  our numerical verification that ideal Bose or Boltzmann  gases under periodic or Neumann boundary conditions exhibit a thermodynamic instability at low temperature near absolute zero.  \\

Although ideal Bose gas has been  studied for decades~\cite{London,Nanda,Ziff,Pethick,Kocharovsky} and discussed in many textbooks~\cite{Huang,Landau,Kadanoff,Reif},  the implication of the constant pressure constraint to the  canonical ensemble of finite $N$  has  been rarely explored.\footnote{The so-called  constant pressure ensemble~\cite{Andersen,Salinas} fixes the pressure and  allows the  volume to fluctuate, as its  partition function  is given by
\[
Y_{N}(\beta,P)=\int{\rm d}V\,e^{-\beta VP}Z_{N}(\beta,V)\,.
\]
However, just like ${C_{V}=(\kB/N)\beta^{2}\partial^{2}_{\beta}\ln Z_{N}}$,  the specific heat therein is positive definite and finite:
\[
0<C_{P}^{\prime}={(\kB/N)}\beta^{2}\partial_{\beta}^{2}\ln Y_{N}<\infty\,.
\]
As we are interested in the precise change of the volume without allowing any  fluctuation of it,
  in the present paper we focus on  the standard canonical ensemble. } To the best of our knowledge    the finite $N$-effect on the canonical ensemble  has been  addressed  only in the case of keeping  the   volume  fixed, rather recently by Kleinert~\cite{KleinertBOOK} and by  Glaum, Kleinert and Pelster~\cite{Glaum}.  The earlier focus was typically on either grand canonical or micro-canonical ensembles, and the computations  often assumed  a continuous approximation to convert discrete sums to integrals~\cite{LondonBook,Chaba},   unless an external harmonic potential sets the sum to be taken   over  a geometric series~\cite{Zho1,Brosens96,Mullin97,Grossmann97,Schmidt98,Schmidt99,Scully99,Kocharovsky00,Holthaus01,MullinFern03}.   \\

%%%
%%In contrast, our scheme is to evaluate  the   thermodynamics of ideal Bose gas in a box exactly,
%%keeping   the pressure constant and  referring only to one exact  canonical partition function of definite $N$.
%%%
%%For the final touch we  shall  employ   a modern  computing power.\footnote{JS20 (PowerPC 970) system.}\newpage
%%%%

\newpage

%%%%%%%%%%%%%%%%%%%%%%%%%%%%%%%%%%%%%%%%%%%%%%%%%%%%%%%%%%%%%%%%%%%%%%%%%%%%%%%%%%%%%
%%%%%%%%%%%%%%%%%%%%%%%%%%%%%%%%%%%%%%%%%%%%%%%%%%%%%%%%%%%%%%%%%%%%%%%%%%%%%%%%%%%%%
%%%%%%%%%%%%%%%%%%%%%%%%%%%%%%%%%%%%%%%%%%%%%%%%%%%%%%%%%%%%%%%%%%%%%%%%%%%%%%%%%%%%%
\section{First-order phase transitions in finite systems\label{sectionGENERAL}}
Here  we explain how   canonical ensembles with finite number of physical degrees may exhibit  a discontinuous  phase transition  featuring a genuine mathematical singularity when we keep not    volume but   pressure fixed.

Prior to  rigorous  analysis, a thought experiment may provide  an intuitive clue for this claim: If we fill a rigid box with water to the full capacity and heat it, the temperature will increase but  hardly  it evaporates. However, once we open  the lid,  simply it boils.

Explicitly,  as  pressure  being a function of $T$ and $V$, from   (\ref{PSEC})  we have
\be
\rmd P=\rmd T\partial_{V}S+\rmd V\partial_{V}P\,.
\ee
Hence under constant pressure the temperature derivative acting on  any function of  $\beta$ and $V$  can be computed as
\be
\textstyle{\left.\frac{\partial~}{\partial T}\right|_{P}=
-\kB\beta^{2}\partial_{\beta}-\left(\partial_{V}S/\partial_{V}P\right)\partial_{V}\,.}
\label{dTP}
\ee
Note that  unless explicitly specified as $\left.\frac{\partial~}{\partial T}\right|_{P}$ and $\left.\frac{\partial~}{\partial T}\right|_{V}$,
throughout the paper   $\partial_{\beta}$  always denotes the beta derivative at  fixed $V$ and  $\partial_{V}$ is the volume derivative at  fixed $\beta$, as already taken in (\ref{PSEC}).

In particular, the specific heat per particle under constant pressure reads
\be
\textstyle{C_{P}=\frac{1}{N}\!\left.\frac{\partial~}{\partial T}\right|_{P}(E+PV)
=C_{V}-\left(\partial_{V}S\right)^{2}/\left(\kB\beta N\partial_{V}P\right)\,.}
\label{CPexpression}
\ee
We pay attention to the denominator here  which is essentially the volume derivative of the pressure:
\be
\partial_{V}P=
\beta\langle\,\left(\partial_{V}E_{\state}-\langle\partial_{V}E_{\state}\rangle\right)^{2\,}
\rangle-\langle\partial^{2}_{V}E_{\state}\rangle\,.
\ee
This quantity possesses an  indefinite sign.  If $\partial_{V}P<0$, the system is stable: it resists against the change of  external pressure by adjusting its volume. On the other hand, the opposite case, $\partial_{V}P\geq 0$,  characterizes the first-order phase transition, as the volume at the phase transition is not single  valued~\cite{Huang}. Clearly from (\ref{dTP}), when $\partial_{V}P=0$,  a singularity  develops and \textit{every} physical quantity should  change discontinuously. Moreover, when $\partial_{V}P$ crosses the vanishing line, $C_{P}$ can be negative.  To the best of our knowledge, all the   known  systems revealing  negative specific heat do not keep the volume constant, such as  in astrophysics~\cite{Thirring,LyndenBell},  melting transitions~\cite{Labastie,GrossD}, and in real experiments~\cite{Agostino,SchmidtNega}.

We emphasize that  from (\ref {dTP})   the only way for a finite canonical ensemble  under constant pressure to reveal singularities is through $\partial_{V}P=0$ \textit{i.e.}  the sign of the thermodynamic instability.\\

We close this section by presenting  a sufficient,  yet unnecessary, condition for the thermodynamic instability.\\

\begin{itemize}
\item \textbf{Theorem}\\
 \textit{At low temperature near  absolute zero, a thermodynamic  system is   unstable \textit{i.e.}~$\partial_{V}P >0$,  if the ground state energy is volume independent.}
\end{itemize}

The proof is simple once we spell the canonical partition function as $Z_{N}=\sum_{n}\Omega_{n}e^{-\beta E_{n}}$, where $E_{n}$'s are the possible energy eigenvalues and $\Omega_{n}$ is the corresponding degeneracy.  By  direct  manipulation we may express   $\partial_{V}P$ in the following form:
\be
\partial_{V}P=(1/\beta)\partial^{2}_{V}\ln Z_{N}=(\Omega_{\fst}/\Omega_{0})e^{-\beta (E_{\fst}-E_{0})}\left[\beta(\partial_{V} E_{\fst})^{2}-\partial_{V}^{2}E_{\fst}~+~\cdots~\right]\,,
\label{PROOF}
\ee
where $E_{0}$ is the volume independent ground state energy, $\partial_{V}E_{0}=0$,  with the degeneracy $\Omega_{0}$; $E_{\fst}$ is the first excited state energy satisfying $\partial_{V}E_{\fst}\neq 0$ with the degeneracy $\Omega_{\fst}$.
The  ellipsis denotes  exponentially suppressed terms for large $\beta$ or low temperature. Clearly  at low temperature near  absolute zero $\partial_{V}P$ becomes positive. This completes our proof.\\

Examples include systems with vanishing ground state energy, such as
supersymmetric models, ideal Bose or Boltzmann gases  under periodic or Neumann boundary conditions.
Appendix contains a numerical verification of the latter.\\

\newpage

%%%%%%%%%%%%%%%%%%%%%%%%%%%%%%%%%%%%%%%%%%%%%%%%%%%%%%%%%%%%%%%%%%%%%%%%%%%%%%%%%%%%%
%%%%%%%%%%%%%%%%%%%%%%%%%%%%%%%%%%%%%%%%%%%%%%%%%%%%%%%%%%%%%%%%%%%%%%%%%%%%%%%%%%%%%
%%%%%%%%%%%%%%%%%%%%%%%%%%%%%%%%%%%%%%%%%%%%%%%%%%%%%%%%%%%%%%%%%%%%%%%%%%%%%%%%%%%%%
\section{Ideal Bose gas confined in a cubic  box\label{sectionIDEAL}}
Based on the general analysis of the previous section, henceforth  as a concrete model we focus on ideal Bose gas  confined in a cubic  box and subject to Dirichlet boundary condition. Since there is no zero mode, the ground state energy depends on the volume and our \textit{theorem} above is  not applicable in this case.  Nevertheless, by numerical analysis we show that  $\partial_{V}P$ therein assumes positive values for some interval of  temperature if $N\geq 7616$.

%%%%%%%%%%%%%%%%%%%%%%%%%%%%%%%%%%%%%%%%%%%%%%%%%%%%%%%%%%%%%%%%%%%%%%%%%%%%%%%%%%%%%
%%%%%%%%%%%%%%%%%%%%%%%%%%%%%%%%%%%%%%%%%%%%%%%%%%%%%%%%%%%%%%%%%%%%%%%%%%%%%%%%%%%%%
%%%%%%%%%%%%%%%%%%%%%%%%%%%%%%%%%%%%%%%%%%%%%%%%%%%%%%%%%%%%%%%%%%%%%%%%%%%%%%%%%%%%%
\subsection{Algebraic analysis}

For non-interacting identical bosonic particle systems, more easily computed than the canonical  partition function is the grand canonical  partition function:
\be
\textstyle{\cZ=\prod_{\vec{n}}\left[
\sum_{j=0}^{\infty}(\eta e^{-\beta E_{\vec{n}}})^{j}\right]=\prod_{\vec{n}}\left(1-\eta e^{-\beta E_{\vec{n}}}\right)^{-1}\,,}
\label{GrandZid}
\ee
where $\eta$ denotes a fugacity and
$\vec{n}$ corresponds to a good quantum number valued ``vector" which uniquely specifies   every quantum state of the \textit{single} particle system.  Taking logarithm and exponentiating  back,  we acquire  an alternative useful expression:
\be
\ba{ll}
\textstyle{\cZ=\exp\!\left(\sum_{k=1}^{\infty}\lambda_{k\,}\eta^{k}/k\right)\,,}~~~&~~
\textstyle{\lambda_{k}:=\sum_{\vec{n}\,}e^{-k\beta E_{\vec{n}}}}\,.
\ea
\label{lambda}
\ee
From the power series expansion of this,  ${\cZ=\sum_{N} Z_{N}\eta^{N}}$,  one can   easily read off the canonical partition function, as previously  obtained by Matsubara~\cite{Matsubara}  and Feynman~\cite{Feynman}:
\be
\textstyle{{Z}_{N}=\sum_{{{m_{a}}}}
\prod_{a=1}^{N}{(\lambda_{a})^{m_{a}}}/{(m_{a}!\,a^{m_{a}})}\,,}
\label{ZNFeynman}
\ee
where the sum  is over all the partitions of   $N$, given by non-negative integers  $m_{a}$, \small{$a=1,2,\cdots, N$} satisfying   $N=\sum_{a=1}^{N} a\,m_{a}$. In particular, with ${\lambda_{1}=Z_{1}}$ the partition as ${m_{1}=N}$  leads to a conventional approximation~\cite{Gibbs}, or the canonical partition function of ideal Boltzmann gas:
 \be
 {Z_{N}\simeq (Z_{1})^{N}/N!}\,.
 \label{conapp}
 \ee
This approximation would be only valid if all the particles  occupied  distinct states, as in the case of  high temperature limit. Other partitions then give  corrections to such  underestimation: Compared to ideal Boltzmann gas, ideal Bose gas has higher probability for the particles to occupy the same quantum state.

Yet,  according to the Hardy-Ramanujan's estimation, the number of  possible partitions  grows  exponentially like $\textstyle{e^{\pi\sqrt{2N/3}}/(4\sqrt{3}N)}$, and this would make any numerical computation  practically hard for large $N$. Alternatively, we consider a recurrence relation on $Z_{N}$,
which was first derived by Landsberg~\cite{Landsberg} and  can be  easily  reproduced   here after differentiating  $\cZ$ in (\ref{lambda}) by  $\eta\,$: With $Z_{0}=1$  we get for $N\geq 1$,
\be
\textstyle{Z_{N}=\left(\sum_{k=1}^{N}\lambda_{k}Z_{N-k}\right)/N\,.}
\label{recZ}
\ee
Further, if we  formally define an ${\infty\times\infty}$ triangularized  matrix $\M$ whose  entries are given by
\be
\M[a,b]{\,:=}\left\{\ba{cl}\lambda_{a{-b}}/(a{-1})~~&~~\mbox{for~~~} a>b\\
0~~&~~\mbox{otherwise}\,,\ea\right.
\ee
the above recurrence relation gets simplified: 
\be
\textstyle{Z_{N}=\sum_{n=0}^{\infty}\,\M[{N{+1}},{n{+1}}]\,Z_{n}}\,,
\ee
such that it has a solution given by  a particular entry of a certain  matrix:
\be
{\textstyle{Z_{N}=({\textstyle{\sum_{k=0}^{\infty} \M^{k}}})[{N{+1}},1]=({I-\M})^{-1}[{N{+1}},1]\,.}}
\label{ZNinv}
\ee
Since $\M^{k}[{N{+1}},1]$ vanishes for   ${k\geq{N{+1}}}$,  the triangularized matrix $\M$   can be effectively - and happily -  truncated to its upper left  ${(N{+1}){\times}(N{+1})}$ finite block.  Then,  from the standard recipe~\cite{Arfken} to compute the inverse of a matrix,\footnote{From ${\det({I-\M}){=}1}$,  an intermediate relation between (\ref{ZNinv}) and (\ref{ZNPsi}) follows:
\[
({I-\M})^{-1}[{N{+1}},1]=(-1)^{N}\det(\Phi_{N})\,,
\]
where  $\Phi_{N}$ is an
$N{\times N}$ matrix whose entry $\Phi_{N}[a,b]$ is given by
$-\lambda_{a{-b}{+1}}{/a}\,$ for $b\leq a$, unity for ${b=a{+1}}$ and zero otherwise.}  another, novel  expression of the canonical partition function follows:
\be
\textstyle{Z_{N}=\det(\Omega_{N})(Z_{1})^{N}/N!}
\label{ZNPsi}
\ee
where $\Omega_{N}$  is an almost triangularized  $N{\times N}$ matrix of which the entries  are  defined  by
\be
\Omega_{N}[a,b]:=\left\{\ba{cl}
\lambda_{a{-b}{+1}}{/\lambda_{1}}~~&~~\mbox{for~~~} {b\leq a}\\
-a{/\lambda_{1}}~~&~~\mbox{for~~~} {b=a+1}\\
0~~&~~\mbox{otherwise}\,.\ea\right.
\ee
In particular, every diagonal entry is unity so that when ${\lambda_{1}\rightarrow\infty}$,  we have
$\det(\Omega_{N})\rightarrow 1$ and hence the reduction: ${Z_{N}\rightarrow(Z_{1})^{N}/N!}$ as in (\ref{conapp}).\\
%%
%%\be
%%\Omega_{N}[a,b]:=\left\{\ba{cl}
%%\lambda_{a{-b}{+1}}{/\lambda_{1}}~&~\mbox{for~~}b<a\\
%%1~&~\mbox{for~~}b=a\\
%%-a{/\lambda_{1}}~&~\mbox{for~~}b=a+1\\
%%0~&~\mbox{otherwise.}
%%\ea\right.
%%\label{Psi}
%%\ee
%%

In addition to the physical quantities (\ref{PSEC}) above,    by considering    $-\beta^{-1}\partial_{E_{\vec{n}}}\ln\cZ=\eta e^{-\beta E_{\vec{n}}}/(1-\eta e^{-\beta E_{\vec{n}}})$ as a trick~\cite{Fujiwara}, we can also compute  the number of particles occupying the ground state:
\be
\textstyle{\langle N_{0}\rangle=\sum_{k=1}^{N}\,e^{-k\beta E_{0}}Z_{N-k}/Z_{N}\,.}
\label{N0value}
\ee
Each term in the sum above corresponds to the probability for at least $k$ particles to occupy the lowest state.  If we denote this probability by $p_{k}$, the difference $p_{k}-p_{k{+}1}$  corresponds to the probability for precisely $k$ particles to occupy the ground state~\cite{Wilkens,HolthausKalin}.  This leads to an alternative derivation of (\ref{N0value})  as: $\langle N_{0}\rangle=\sum_{k=1}^{N} k(p_{k}-p_{k{+}1})=\sum_{k=1}^{N} p_{k}~$ where $~p_{{N}{+}1}=0$.\\

Henceforth, exclusively  for ideal Bose gas we focus on $N$  particles with mass $m$, confined  in a  box  of dimension $d$ and length $L\equiv V^{1/d}$.   Hard, impenetrable walls impose Dirichlet boundary condition~\cite{HolthausKS}. Since we are interested in a finite system,  the  periodic  boundary condition which is somewhat more popular in the literature is  not suitable for our purpose. We recall that nevertheless enforcing  periodic or Neumann boundary condition leads to a thermodynamic instability at low temperature near absolute zero for arbitrary $N$  (see Figure 3 in Appendix).

With  \textit{positive} integer valued good quantum numbers:
\be
\vec{n}=(n_{1},n_{2},\cdots,n_{d\,})\,,
\ee
the  single particle Boltzmann factor in (\ref{GrandZid}) assumes the form:
\be
\ba{ll}
e^{-\beta E_{\vec{n}}}=q^{\vec{n}{\cdot\vec{n}}}\,,~~&~~\displaystyle{
q:=e^{-\beta\pi^{2}\hbar^{2}/(2m  V^{2/d})}\,.}
\ea
\label{Boltzmann}
\ee
In terms of a Jacobi  theta function:
\be
\textstyle{\vartheta(q):=[\theta_{3}(0,q)-1]/2=\sum_{n=1}^{\infty} q^{n^{2}}\,,}
\label{vartheta}
\ee
we get specifically for (\ref{lambda}):
\be
\lambda_{k}(q)=[\vartheta(q^{k})]^{d}\,.
\ee

\noindent After all, {${Z}_{N}$ becomes  a function of only one  variable  $q$}, so that  we may  put
\be
\ba{ll}
\beta\partial_{\beta}\equiv q\ln q\partial_{q}\,,~~~~&~~~~
V\partial_{V}\equiv -(2/d)q\ln q\partial_{q}\,.
\ea
\ee
This implies that all the dimensionless physical quantities such as $C_{V}/\kB$, $C_{P}/\kB$, $\langle N_{0}\rangle$, \textit{etc.} are also functions of the single variable $q$. Consequently, it turns out that the temperature dependence of all these  dimensionless quantities can be best analyzed  if we introduce   the following two  dimensionless ``temperatures":
\be
\ba{l}
\tau_{V}:=\kB T (V/N)^{2/d}(2m/\left(\pi^{2}\hbar^{2})\right)=(4/\pi)(\leff/\LdeB)^{2}\,,\\
\tau_{P}:=\kB T P^{-2/(d+2)}\left(2m/(\pi^{2}\hbar^{2})\right)^{d/(d+2)}\,,\\
\ea
\label{tauVP}
\ee
where  $\LdeB=\hbar\sqrt{2\pi/(m\kB T)}$ is the thermal de Broglie wavelength and $\leff=(V/N)^{1/d}$ is the average interparticle distance.  While the former in (\ref{tauVP}) is a monotonically increasing function of $q$,  the latter may be not so as:
\be
\ba{ll}
-1/\tau_{V}=N^{2/d}\ln q\,,~~~~&~~~~
 -1/\tau_{P}=\ln q[(2/d)q\partial_{q}\ln Z_{N}]^{2/(d+2)}\,.
 \ea
 \ee
Any critical value of these quantities will  automatically give us  the critical temperature at an  arbitrarily given   volume or  pressure.

In a similar fashion,  we  also define   a dimensionless indicator of the thermodynamic instability,    $\phi$, as well as dimensionless  ``volume" and  ``energy":
\be
\ba{l}
\,\phi\,\,:=-(1/N)\beta V^{2}\partial_{V}P=4C_{V}^{2}/[d^{2}\kB(C_{P}-C_{V})]\,,\\
\upsilon_{P}:=(\tau_{V}/\tau_{P})^{d/2}=(V/N)\left(2mP/(\pi^{2}\hbar^{2})\right)^{d/(d+2)}\,,\\
\epsilon_{P}:=(E/N)P^{-2/(d+2)}\left(2m/(\pi^{2}\hbar^{2})\right)^{d/(d+2)}\,.
\ea
\label{phi}
\ee
\newpage

\noindent $\bullet$ \textit{Low temperature limit}: \\
The variable $q$ lies between zero and one.
As $q\rightarrow 0$ we have
\be
Z_{N}\rightarrow e^{-\beta NE_{0}}=q^{Nd}\,.
\label{Zatzero}
\ee
Hence at $q=0$, the temperatures vanish  $\tau_{V}=\tau_{P}=0$, and
\be
\ba{llll}
{C_{V}=C_{P}=0}\,,~&~{\phi=\infty}\,,~&~{\langle N_{0}\rangle=N}\,,~&~
{\upsilon_{P}=(2^{d}/N^{2})^{1/(d+2)}}\,.
\ea
\label{qzero}
\ee
In particular, the volume reads at absolute zero,
\be
V=[N\pi^{2}\hbar^{2}/(mP)]^{d/(d+2)}\,.
\label{Vzero}
\ee
That is to say, despite of  the apparent Bose-Einstein condensation \textit{i.e.~}${\langle N_{0}\rangle=N}$, the  volume assumes a finite value which is even not  extensive. The finiteness is  essentially due to the Heisenberg uncertainty principle: Since  the particles are localized in a finite box, the  uncertainty principle forbids the ground state energy $E_{0}$ to vanish  and leads to the nontrivial canonical partition function~(\ref{Zatzero}).\\

\noindent $\bullet$ \textit{High temperature limit}:\\
As $q\rightarrow 1$,  from (\ref{ZNPsi}) and thanks to an identity of the theta function by Jacobi~\cite{Jacobi}:
\be
\vartheta(e^{-\pi\sigma})+1/2=[\vartheta(e^{-\pi/\sigma})+1/2]/\sqrt{\sigma}\,,
\ee
we have
\be
Z_{N}\rightarrow (1/N!)(-\pi/4\ln q)^{Nd/2}\,.
\ee
 Hence at $q=1$, the temperatures diverge
$\tau_{V}=\tau_{P}=\infty$, and
\be
\ba{lll}
{C_{V}/\kB=d/2\,,}~&~{C_{P}/\kB=1+d/2\,,}~&~{\phi=1\,,}\\
{\upsilon_{P}/\tau_{P}=1\,,}~&~{(\tau_{V})^{d/(d+2)}/\tau_{P}=1\,,}~&~{\langle N_{0}\rangle=0\,.}
\ea
\label{qone}
\ee
Namely,   ideal Bose gas reduces to the  classical ideal gas at high temperature, satisfying the classic  relation:
\be
PV=N\kB T\,.
\label{classicideal}
\ee
\newpage

%%%%%%%%%%%%%%%%%%%%%%%%%%%%%%%%%%%%%%%%%%%%%%%%%%%%%%%%%%%%%%%%%%%%%%%%%%%%%%%%%%%%%
%%%%%%%%%%%%%%%%%%%%%%%%%%%%%%%%%%%%%%%%%%%%%%%%%%%%%%%%%%%%%%%%%%%%%%%%%%%%%%%%%%%%%
%%%%%%%%%%%%%%%%%%%%%%%%%%%%%%%%%%%%%%%%%%%%%%%%%%%%%%%%%%%%%%%%%%%%%%%%%%%%%%%%%%%%%
\subsection{Numerical results\label{sectionNUMERICAL}}
Here we present our numerical results of $d=3$, \textit{i.e.~}cubic box, at generic temperature.
 Our  analysis  is based on a set of  recurrence relations which  enables us to perform $N^{2}$ order computation, enhanced by a parallel computing   power - JS20 (PowerPC 970) system.
Essentially we utilize  (\ref{recZ})\footnote{Since $Z_{N}$ can be a big number for large $N$, for the evaluation of the canonical partition function itself, it is convenient  to decompose it as
\[
\textstyle{Z_{N}=\prod_{n=1}^{N}\,f_{n}}\,,
\]
and  utilize  a recurrence  relation:
\[
\textstyle{f_{N}=\sum_{n=0}^{N-1}\,[\lambda_{N{-n}}/(N\prod_{j={n+}1}^{N{-1}}f_{j})]}\,,
\]
which  is equivalent to (\ref{recZ}).} and two other relations which can be  straightforwardly obtained from the grand canonical partition function expressed  in the form (\ref{lambda}):
\be
\ba{l}
\displaystyle{\beta\partial_{\beta}Z_{N}=\sum_{n=1}^{N}\,\rho_{n}\lambda_{n}Z_{N{-n}}\,,}\\
\displaystyle{\beta^{2}\partial^{2}_{\beta}Z_{N}=
\sum_{n=1}^{N}\,[n(\zeta_{n}+\rho_{n}^{2})\lambda_{n}Z_{N{-n}}+
\rho_{n}\lambda_{n}\,\beta\partial_{\beta}Z_{N{-n}}]\,,}
\ea
\ee
where we set
\be
\ba{ll}
{\rho_{n}:=(\beta/n)\partial_{\beta}\ln\lambda_{n}}\,,~~~~&~~~~
{\zeta_{n}:=(\beta^{2}/n^{2})\partial^{2}_{\beta}\ln\lambda_{n}}\,.
\ea
\ee
All our results agree with   the asymptotic behaviours  (\ref{qzero}), (\ref{qone}).\\

\noindent$\bullet$ \textit{Constant volume curves} (Figure 1):\\
As expected, all the physical quantities are smooth single valued functions of the temperature $\tau_{V}$. As  $N$ grows, the specific heat $C_{V}$ develops a maximum, at which $\langle N_{0}\rangle$ drops rapidly. This behaviour is consistent  with Ref.\cite{Glaum}.  More importantly for us,
while $\phi$ decreases from infinity at $\tau_{V}=0$ to one at $\tau_{V}=\infty$, it develops a local minimum which becomes eventually  negative  if  $N\geq 7616$. This manifests  the thermodynamic instability of the ideal Bose gas confined in a cubic box.\newpage
\noindent
\begin{figure}[H]
\normalsize
\includegraphics[width=75mm]{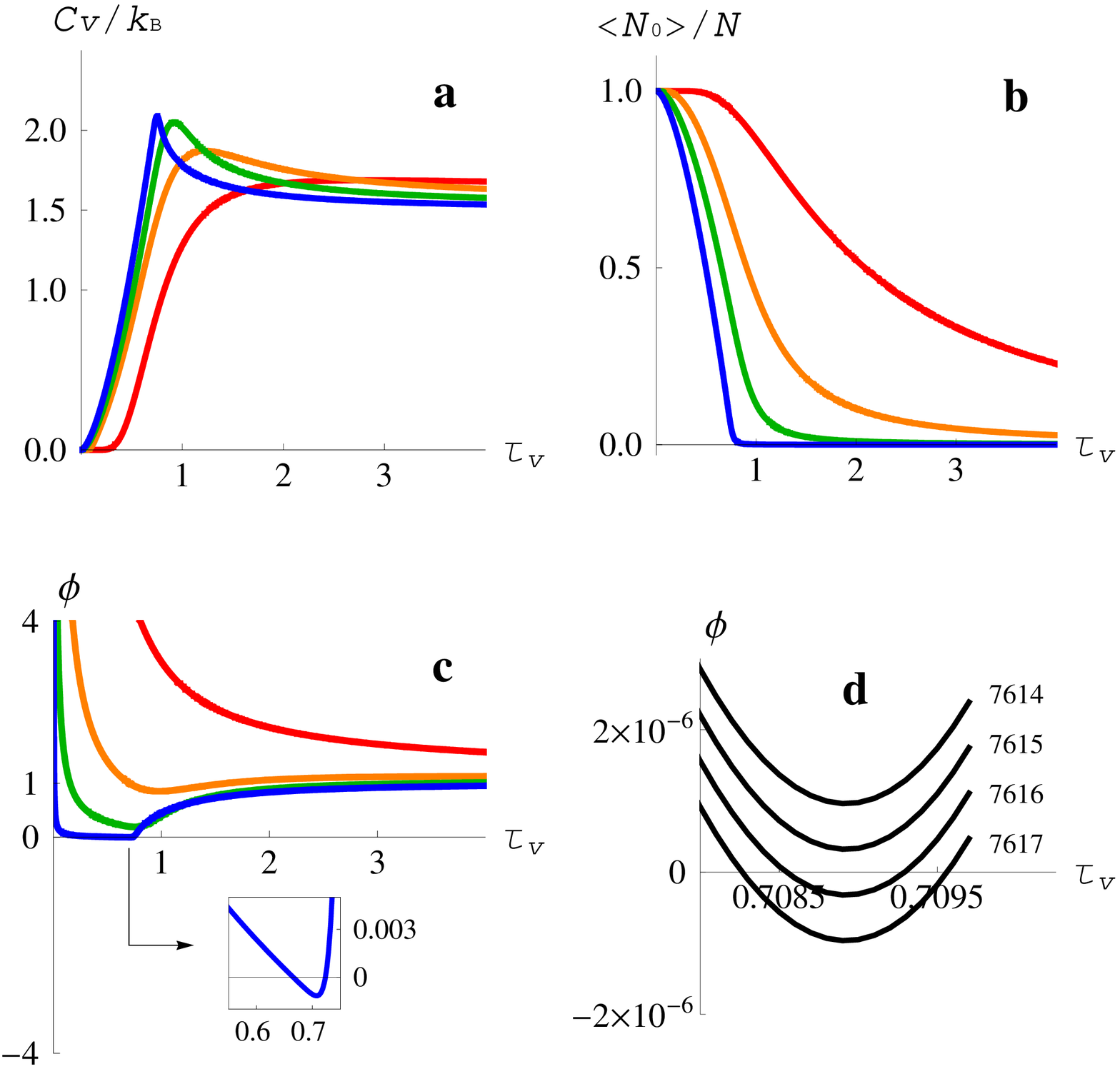}
\caption[]{{\textbf{Constant volume curves}} \\
{\bf a:}   specific heat per particle at constant volume, $C_{V}/\kB$\\
{\bf b:} the occupancy ratio of  the ground state, $\langle N_{0}\rangle/N$\\
{\bf c,\,d:} the thermodynamic   instability indicator, $\phi=-(1/N)\beta V^{2}\partial_{V}P$\\
Here {\textbf{a,b,c}} are  for  $N{=\red{1}}$ (red), $\orange{10}$ (orange), $\green{10^{2}}$ (green), $\blue{10^{4}}$ (blue), while {\textbf{d}} is   for \textbf{ $N=7614,7615,7616,7617$}. The case of $N{=\red{1}}$ (red)  also  corresponds to ideal Boltzmann gas. All the quantities are dimensionless. }
\label{Fig1for7616}
\end{figure}

\newpage

\noindent
\begin{figure}[H]
\normalsize
\includegraphics[width=75mm]{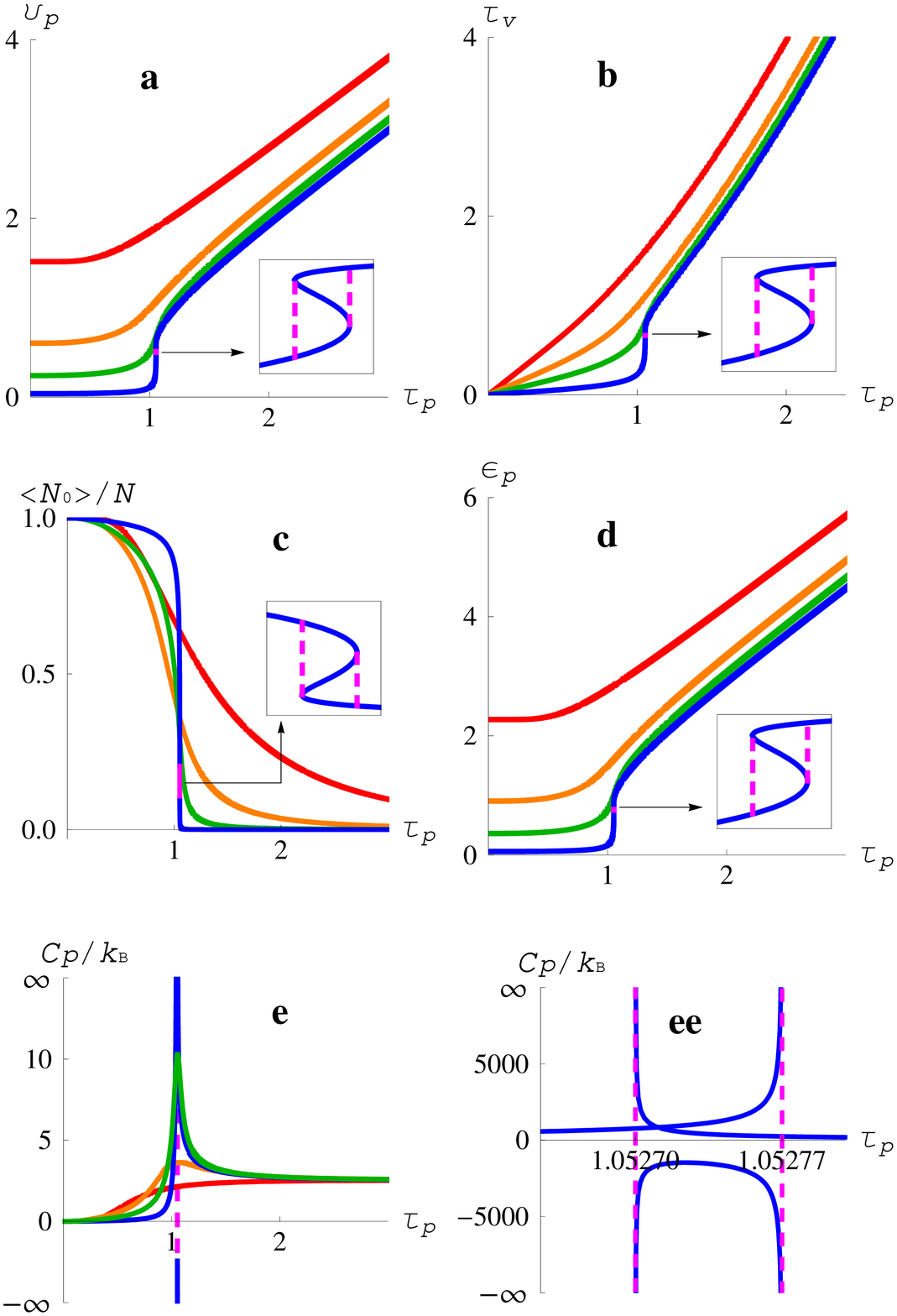}
\caption[]{{\textbf{Constant pressure curves}} \\
{\bf a:} the dimensionless volume $\upsilon_{P}$ versus the dimensionless temperature $\tau_{P}$\\
{\bf b:} $\tau_{V}$ versus $\tau_{P}$\\
{\bf c:} $\langle N_{0}\rangle/N$ versus $\tau_{P}$\\
{\bf d:} the dimensionless energy $\epsilon_{P}$ versus $\tau_{P}$\\
{\bf e,\,ee:}   specific heat per particle under constant pressure  $C_{P}/\kB$ versus $\tau_{P}$\\
Note that {\textbf{a,b,c,d,e}} are  for \textbf{ $N{=\red{1}}$} (red) \textit{i.e.~}ideal Boltzmann gas  or\\  $N{=\orange{10}}$ (orange), $\green{10^{2}}$ (green), $\blue{10^{4}}$ (blue).
The small boxes magnify the zigzag segments  of \textbf{$N=\blue{10^{4}}$} (blue), while {\textbf{{ee}}} magnifies {\textbf{{e}}} for   \textbf{$N=\blue{10^{4}}$} (blue). The dotted pink lines denote  supercooling and superheating points. }
\label{Fig2for7616}
\end{figure}
\newpage

\noindent$\bullet$ \textit{Constant pressure curves} (Figure 2):\\
Since the thermodynamic  instability indicator $\phi$ vanishes at two points when  $N\geq 7616$,  there are generically two critical temperatures,
$\tau_{P}^{\ast}<\tau_{P}^{\ast\ast}$, which we  suggest to identify  as supercooling and superheating points respectively.  As predicted  from the general argument (\ref {dTP}),  all the physical quantities change discontinuously at these points: On  each physical quantity versus temperature   plane, the  constant pressure line zigzags between  the two  critical points  keeping them as two turning points. Accordingly,  physical quantities are triple valued between the two   points  and double valued at the points. This implies the existence  of  three  different phases   during the phase transition, say two `forward' phases and one `backward' phase.

Specifically, the volume expands abruptly (Figure 2{\bf a}): below the supercooling point it is almost constant whilst  beyond  the superheating point it follows  the classical  ideal gas law (\ref{classicideal}), which resembles the usual liquid-gas transition.  At the same time, the number of particles on the ground state drops abruptly   from the full occupancy to the total evacuation (Figure 2{\bf c}). This  clearly  realizes  a Bose-Einstein condensation taking place  both in the momentum  and in the  coordinate space~\cite{Lamb}.
 The specific heat per particle under constant pressure diverges at the critical points, and when magnified between  the supercooling and the superheating  points (Figure 2{\bf ee}),  intriguingly  it reveals one  negative heat capacity  in addition to other two   positive ones.  Because
${C_{P}/\kB}={{{\rm d}\epsilon_{P}}/{{\rm d}\tau_{P}}}+{{{\rm d}\upsilon_{P}}/{{\rm d}\tau_{P}}}$,
%%%
%%$NC_{P}=\left.\frac{\partial E}{\partial T}\right|_{P}+P\left.\frac{\partial V}{\partial T}\right|_{P}$,
%%%
discontinuous  changes in  both the internal energy and the volume  simultaneously contribute to
the divergence of the specific heat.\footnote{In fact, the  exponent corresponding to  the singularities can be shown to be  $1/2$ \cite{relativistic7616}. }

Algebraically, while  physical quantities are \textit{a priori} single valued  functions of $q$ and hence $\tau_{V}$,  the origin of the  multi-valuedness and  singular behaviours   can be all  traced back to the  zigzagging of the isobar  curve on $(\tau_{V},\tau_{P})$ plane  (Figure 2{\bf b}). Namely the isobar curves of the ideal Bose gas zigzag  on the $(V,T)$ plane, if $N\geq 7616$.

Our numerical results  of the supercooling and the superheating  points are listed below for  selected  $N$:
%\noindent
%\begin{table}[H]
\begin{center}
\begin{tabular}{c|l|l}
%\hline
{$N$}&$~~~~~~\tau_{P}^{\ast}$  &~~~~~$\tau_{P}^{\ast\ast}$  \\
\hline
$7616~~$&$~~1.0543694113~~$&$~~1.0543694116~~$\\
$10^{4}\,$&$~~1.05270~~$&$~~1.05277~~$\\
$10^{5}\,$&$~~1.0410~~$&$~~1.0424~~$\\
$10^{6}\,$&$~~1.034~~$&$~~1.036~~$
%\hline
\end{tabular}
%\caption{For different $N$  supercooling and superheating  points are }
\end{center}
%\label{TheTable}
%\end{table}
~\newline
from which, recovering all the dimensionful parameters, we may read off the supercooling and the superheating temperatures  at arbitrary constant pressure for  each value of $N$:
\be
\ba{l}
\kB T_{\superc}\simeq \tau_{P}^{\ast}\times\textstyle{(\frac{\pi^{2}\hbar^{2}}{2m})^{3/5}P^{2/5}}\,,\\
\kB T_{\superh}\simeq \tau_{P}^{\ast\ast}\times\textstyle{(\frac{\pi^{2}\hbar^{2}}{2m})^{3/5}P^{2/5}}\,.
\ea
\label{superch}
\ee

With the mass of Helium-4 and under the pressure $1$ atm, our numerical result of $N=10^{6}$ gives us the supercooling temperature $1.686$ Kelvin  and the superheating temperature $1.689$ Kelvin, both of which are of the same order as   the lambda point $2.17$ Kelvin or alternatively as the boiling point $4.22$ Kelvin.\newline

%%%%%%%%%%%%%%%%%%%%%%%%%%%%%%%%%%%%%%%%%%%%%%%%%%%%%%%%%%%%%%%%%%%%%%%%%%%%%%%%%%%%%%%%%%%%%%%%%
%%%%%%%%%%%%%%%%%%%%%%%%%%%%%%%%%%%%%%%%%%%%%%%%%%%%%%%%%%%%%%%%%%%%%%%%%%%%%%%%%%%%%%%%%%%%%%%
%%%%%%%%%%%%%%%%%%%%%%%%%%%%%%%%%%%%%%%%%%%%%%%%%%%%%%%%%%%%%%%%%%%%%%%%%%%%%%%
\section{Discussion\label{sectionDISCUSSION}}
In summary,    finite amount of  ideal Bose gas confined in a cubic box  reveals a thermodynamic instability, if  $N\geq 7616$. This implies that, under  constant  pressure condition, the ideal Bose gas may undergo    a first-order phase transition accompanied  by  genuine mathematical singularities. It  is characterized by  both the supercooling and the  superheating points; Bose-Einstein condensation in both the momentum and the coordinate spaces; and the very fact that all the physical quantities become  triple valued between the two points. In particular, one of the three values of the  specific heat under constant pressure  is negative.   While Bose-Einstein condensation  appears as a  continuous  phase transition when the volume is kept constant  (Figure 1{\,\textbf{b}}),  it may become  a discrete phase transition   if the pressure is held fixed (Figure 2{\,\textbf{c}}).

Ideal gas consists of  featureless point particles. It is such a  simple model that the canonical partition function becomes essentially a one-variable function, $Z_{N}(q)$.  Assuming an extra internal structure of the particle, \textit{e.g.~}spin or a vibrational mode, will  introduce an additional dimensionful   parameter into the energy spectrum.
This may  generate an additional (continuous or discontinuous)   phase transition and distinguish  the Bose-Einstein condensation to the ground state from  the condensation in the coordinate space~\cite{progress}.

The critical number 7616 we report  in this paper   can be regarded as a characteristic number of `cube' that is the geometric shape of the box containing  the ideal Bose gas.  Boxes of different shapes (see \textit{e.g.~}\cite{cuboid})  will have  different critical numbers.  Thus,  our scheme of investigating  the thermodynamic instability   of  ideal Bose gas can provide a novel  algorithm to assign a characteristic number to each geometric closed two-dimensional manifold.

Apparently ideal Bose gas has no  interaction. However, compared to ideal Boltzmann gas, ideal Bose gas has higher probability for the particles to occupy the same quantum state, as  seen from  the comparison between (\ref{ZNFeynman}) and (\ref{conapp}).\footnote{As a simple example, consider  a two-particle system with quantum states `up' and `down' only. The probability for the  two identical bosonic particles to occupy the same  `up' state is $1/3$ and this is higher than that for the case of two Boltzmann particles, $1/4$. Similarly, the probability for the  two identical  bosonic particles to occupy the two different   states is $1/3$ which is  lower than that  of the two Boltzmann particles, $1/2$.}  In coordinate space this means  that identical bosonic particles have tendency to  gather together in comparison   to  ideal Boltzmann gas.   In fact, a path integral representation of the  canonical partition function of ideal Bose gas reveals an existence of an   attractive effective  potential~\cite{Huang}.\footnote{It is worth while to note that,   the nontrivial volume at absolute zero due to the Heisenberg uncertainty principle  (\ref{Vzero}) suggests a statistical  repulsive interaction at  low temperature. Nevertheless  this  is  valid  for ideal Boltzmann gas too. For related subtle issues see \textit{e.g.} \cite{statisticalforce}.}   This  provides a physical  clue   to  the condensation in both the momentum and the coordinate spaces: As the temperature decreases,  the effective  statistical   attraction becomes dominant  and the system condensates. In this context,  it  is also worth while   to recall  the similarity between the permutation symmetry of the identical particle indices and the gauge symmetry of the standard model  in high energy physics  or  matrix models (see \textit{e.g.}~\cite{KraussWilczek89}). Although the former is discrete while the latter is continuous,  the latter may include the former as  a subgroup. The common feature is that   they both correspond to \textit{nonphysical} symmetry~\cite{Park}. As a matter of  fact, up to an overall factor, the canonical  partition function of  identical harmonic oscillators~\cite{Zho1}  coincides with that of a massive Yang-Mills quantum mechanics~\cite{Z10matrix}, taking  the form $\prod_{n=1}^{N} (1-q^{n})^{-1}$.

Generically, for a stable matter  $\partial_{V}P$ is negative.   What we show by taking  ideal Bose gas as an exactly solvable model  is  an explicit demonstration   that, if there are sufficiently,  yet finitely, many  identical bosonic particles, $\partial_{V}P$ can be positive for at least one  interval of temperature.
It will be therefore interesting and crucial  to see, to what extent   interactions  can alter this. If not much,  as in a weakly interacting system,    one first-order phase transition accompanying  a discontinuous  volume change, or the liquid-gas transition itself,  is likely to occur  essentially due to the identical nature of   particles.
In a perturbative  analysis of the interaction,  ideal Bose gas provides  the `zeroth' order contribution: For each quantum state $\psi$, as the energy eigenvalue is shifted from $E_{\psi}$ to $E_{\psi}+\Delta_{\psi}$, the canonical partition function changes from
$Z_{N}$ to $Z_{N}\langle e^{-\beta\Delta_{\psi}}\rangle$.  Thus,   the thermodynamic instability is modified as
\[
\phi~~\longrightarrow~~\phi\,-\,(1/N)V^{2}\partial^{2}_{V}\ln\langle e^{-\beta\Delta_{\psi}}\rangle\,=\,\phi\,+\,(1/N)\beta V^{2}\partial^{2}_{V}\langle\Delta_{\psi}\rangle\,+\,\cdots\,.
\]
If the first order correction gives a negative contribution, $\partial^{2}_{V}\langle\Delta_{\psi}\rangle\leq 0$, the thermodynamic instability persists naturally  up to the order.
In this case,  the average bonding energy between particles will also  change discontinuously during the phase transition,  but this  will be  an accompanying  side effect rather than  a key reason to drive  the phase transition.

It will be experimentally challenging to find a corresponding critical number for each molecule to manifest a discontinuous phase transition or its  liquid-gas transition under constant pressure. A criterion   for the first-order phase transition     is to observe the supercooling and the superheating phenomena.   \\
~\\
~\\
~\\
\noindent\textbf{Acknowledgements}\\
 We  thank  Imtak Jeon, Jeenu Kim and  In-Ho Lee for helpful comments.  
The work is supported   
%%%
%%by  the Korea Foundation for International Cooperation of Science \& Technology with grant number K20821000003-08B1200-00310,
%%%
by two National Research Foundation of Korea (NRF) grants, 2005-0049409 and 2009-0083765,  funded by the Korea government (MEST). %, through the Center for Quantum Spacetime (CQUeST) of Sogang University. 
\newpage

\appendix

\section*{Appendix: Other boundary conditions}
Here as an explicit demonstration  of the \textit{theorem} (\ref{PROOF}), we present  numerical results (Figure 3)  on the ideal Boltzmann gas (or equivalently  ideal Bose gas with $N{=1}$)  in a cubic box under   boundary condition: Neumann or  periodic. This corresponds to taking $\lambda_{k}(q)=[1+\vartheta(q^{k})]^{d}$ for the former and  
$\lambda_{k}(q)=[1+2\vartheta(q^{4k})]^{d}$  for the latter. Actually we only need $\lambda_{1}(q)$ and  set  $d=3$ as for the cubic box.

 As expected, due to the existing zero mode, both boundary conditions lead to a thermodynamic instability \textit{i.e.} 
 $\partial_{V} P>0$ at low temperature near absolute zero.

\noindent
\begin{figure}[H]
\normalsize
\includegraphics[width=110mm]{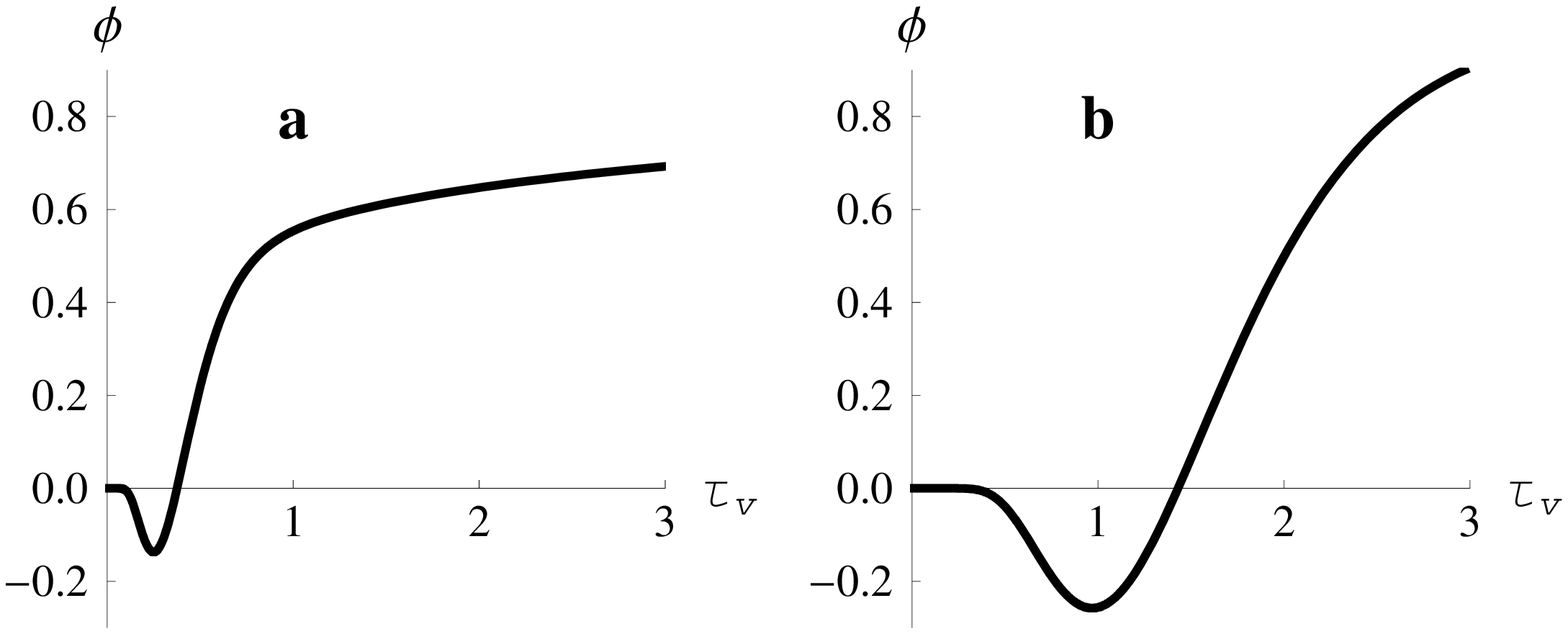}
\caption[]{{\textbf{Thermodynamic instability due to zero mode}} \\
{\bf a:} $\phi$ versus $\tau_{V}$ under Neumann boundary condition\\
{\bf b:} $\phi$ versus $\tau_{V}$ under periodic boundary condition}
\label{Fig7616bc1}
\end{figure}

%%%%%%%
%%\textbf{Author Contributions}  J.-H.P. designed the research, derived the formulae and wrote
%%the paper.  S.-W.K.  checked the formulae, wrote Fortran codes, conducted the numerical
%%analysis and commented  on the manuscript.\\
%%%%%%%
%%\textbf{Author Information} The authors declare
%%that they have no competing financial interests.
%%Correspondence and requests for materials should be addressed to J.-H. Park %%(\,park@sogang.ac.kr\,).%%%\clearpage
%%%%{}\\
%%%%\textbf{Supplementary Information}(SI) is provided.
%%

\end{document}